\documentclass[aps,twocolumn,showpacs,pra,twoside,amssymb,amsmath,latexsym]{revtex4}
\usepackage{graphicx}

\def\cB{{\cal B}}

\def\fid{F}

\def\ket#1{|#1\rangle}
\def\bra#1{\langle#1|}

\begin{document}
\title{Hypothesis testing for an entangled state produced\\ 
by spontaneous parametric down conversion}
\author{Masahito Hayashi$^{\ast,\dag}$,
Bao-Sen Shi$^{\ast}$,
Akihisa Tomita$^{\ast,\ddag}$,\\
Keiji Matsumoto$^{\ast,\S}$,
Yoshiyuki Tsuda$^{\P}$,
and
Yun-Kun Jiang$^{\ast}$}
\affiliation{$^{\ast}$ERATO Quantum Computation and Information Project, 
Japan Science and Technology Agency (JST), Tokyo 113-0033, Japan \\
$^{\dag}$Superrobust Computation Project,
Information Science and Technology Strategic Core (21st Century COE by 
MEXT)\\
Graduate School of Information Science and Technology,
The University of Tokyo\\
7-3-1, Hongo, Bunkyo-ku, Tokyo, 113-0033, Japan\\
$^{\ddag}$Fundamental Research Laboratories, NEC, Tsukuba 305-8501, Japan\\
$^{\S}$National Institute of Informatics, Hitotsubashi, Chiyoda-ku, Tokyo 101-8430, Japan\\
$^{\P}$COE, Chuo University, Kasuga, Bunkyo-ku, Tokyo 112-8551, Japan}
\pacs{03.65.Wj,42.50.-p,03.65.Ud}
%02.20.-a 	Group theory 
%(for algebraic methods in quantum mechanics, see 03.65.Fd; 
%for symmetries in elementary particle physics, see 11.30.-j)
%03.65.Ta Foundations of quantum mechanics; measurement theory 
%(for optical tests of quantum theory, see 42.50.Xa)
%03.65.Ud Entanglement and quantum nonlocality 
%(e.g. EPR paradox, Bell's inequalities, GHZ states, etc.) 
%(for entanglement production in quantum information, see 03.67.Mn; 
%for entanglement in Bose-Einstein condensates, see 03.75.Gg)
%03.65.Wj State reconstruction, quantum tomography
%42.50.-p 	Quantum optics
%\date{\today}
\begin{abstract}
Generation and characterization of entanglement are crucial tasks 
in quantum information processing. 
A hypothesis testing scheme for entanglement has been formulated. 
Three designs were proposed to test the entangled photon states 
created by the spontaneous parametric down conversion.
The time allocations between the measurement vectors were designed to consider
the anisotropic deviation of the generated photon states 
from the maximally entangled states. 
The designs were evaluated in terms of the p-value based on the observed data.
It has been experimentally demonstrated
that the optimal time allocation between the coincidence
and anti-coincidence measurement
vectors improves the entanglement test. 
A further improvement is also 
experimentally 
demonstrated by 
optimizing the time allocation between the anti-coincidence vectors.
Analysis on the data obtained in the experiment verified the advantage 
of the entanglement test designed by the optimal time allocation.
\end{abstract}
\maketitle
%\tableofcontents

\section{Introduction}
The concept of entanglement has been thought to be the heart of 
quantum mechanics. 
The seminal experiment by Aspect et al. \cite{APG82} has proved the 'spooky' 
non-local action of quantum mechanics by observing violation of Bell inequality
 \cite{Bell93} with entangled photon pairs. 
Recently, entanglement has been also recognized as an important resource 
for quantum information processing, 
explicitly or implicitly.  
For example, entanglement provides an exponential speed-up in some computational tasks \cite{Shor97}, and unconditional security in cryptographic 
communications \cite{SP00}. 
A hidden entanglement between the legitimate parties guarantees the security 
in BB84 quantum cryptographic protocol \cite{SP00, BB84}.
Quantum communication between arbitrarily distant parties has been shown to be
possible by a quantum repeater \cite{Briegel98} based on quantum 
teleportation \cite{Bennett93}. 
Practical realization of entangled states is therefore one of the most important issues in the quantum information technology. 
  
The practical implementation raises a problem to verify the amount of 
entanglement.
A quantum information protocol requires a minimum entanglement.
It is, however, not always satisfied in actual experiments.
Unavoidable imperfections will limit the entanglement in generation process.
Moreover, decoherence and dissipation due to the coupling with the environment 
will degrade the entanglement during the processing. 
Therefore, it is a crucial issue to characterize the entanglement 
of the generated (or stored) states to guarantee 
the successful quantum information processing.
For this purpose, quantum state estimation and Quantum state tomography 
are known as a method of identifying the unknown state 
\cite{Selected,helstrom,holevo}.
Quantum state tomography  \cite{WJEK99} has recently applied to obtain full
 information of the $4 \times 4$ two-particle density matrix 
 from the coincidence counts of 16 combinations of measurement \cite{usami}. 
However, characterization is not the goal of an experiment, 
but only a part of preparation. 
It is thus favorable to reduce the time for characterization and 
the number of consumed particles as possible. 
An entanglement test, which should be simpler than the full characterization,
 works well in most applications, 
because we only need to know whether the states are sufficiently entangled 
or not. 
We can reduce the resources for characterization with the entanglement test. 
Barbieri et al. \cite{BMNMDM03} introduced an entanglement witness 
to test the entanglement of polarized entangled photon pairs. 
Tsuda et al. \cite{TMH05} studied the optimization problem 
on the entanglement tests with the mathematical statistics 
in the POVM framework.
Hayashi et al. \cite{HTM} treated this optimization problem
in the framework of Poisson distribution,
which describes the stochastic behavior of the measurement outcomes on
the two-photon pairs generated by spontaneous parametric down conversion
 (SPDC). 

In this paper, we will apply the experimental designs proposed in 
ref. \cite{HTM}
to test the polarization entangled two-photon pairs generated 
by SPDC. The two-photon states can be characterized by the correlation 
of photon detection events in several measurement bases. 
In experiments, the correlation is measured by coincidence counts 
 of photon detections
on selected polarizations. 
The coincidence counts on a combination of the two from
horizontal ($H$), vertical ($V$), $45^{\circ}$ linear ($X$), $135^{\circ}$ 
linear ($D$), clock-wise circular ($R$), and anti-clock-wise circular ($L$) 
polarizations defines a measurement vector.  
We here use the same representations for the measurement vectors as
the state vectors, such as $\vert HH \rangle$.
 If the state is close to 
$\vert \Phi^{(+)} \rangle=\frac{1}{\sqrt{2}}(\vert HH\rangle+\vert VV\rangle)$, the coincidence counts on the vectors $\vert HH\rangle, \vert VV\rangle,
\vert DD\rangle, \vert XX\rangle, \vert RL\rangle,$ and $\vert LR\rangle $ 
yield the maximum values, whereas the counts on the vectors
$\vert HV\rangle, \vert VH\rangle, \vert DX\rangle, \vert XD\rangle, 
\vert RR\rangle,$ and $\vert LL\rangle $ take the minimum values. 
We will refer to the former vectors 
as the coincidence vectors, 
and to the latter as the anti-coincidence vectors. 
The ratio of the minimum counts to the maximum counts measures the degree of
 entanglement.
In the following sections, we formulate the hypothesis testing of entanglement 
in the view of statistics. 
We then improve the test by optimizing the allocation 
on the measurement time for each measurement, 
considering that the counts on the anti-coincidence vectors
are much smaller than those on the coincidence vectors. 

The test can be further improved, if we utilize the knowledge on the tendency
 of the entanglement degradation. 
In general, the error from the maximally entangled states can be anisotropic,
which reflects the generation process of the states. 
We can improve the sensitivity to the entanglement degradation 
by focusing the measurement on the expected error directions.
In the present experiment, we generated polarization entangled photon pairs 
from a stack of two type-I phase matched nonlinear crystals 
\cite{KWWAE99,NUTMK02}, where
one nonlinear crystal generates a photon pair polarized 
in the horizontal direction $(\vert HH \rangle)$, 
and the other generates a photon pair polarized in the vertical direction 
$(\vert VV \rangle)$.
If the two pairs are indistinguishable, the generated photons are entangled 
in a two-photon state 
$\frac{1}{\sqrt{2}} (\vert HH \rangle+ \exp[i \delta] \vert VV \rangle)$. 
Otherwise, the state will be a mixture of $HH$ pairs and $VV$ pairs. 
The quantum state tomography has shown that the only $HHHH$, $VVVV$, 
$VVHH$, $HHVV$ elements are dominant \cite{NUTMK02}, which implies 
that the density matrix can be approximated by a classical mixture 
of $\vert \Phi^{(+)} \rangle \langle \Phi^{(+)} \vert$ and 
$\vert \Phi^{(-)} \rangle \langle \Phi^{(-)} \vert$.
 We can improve the entanglement test on the basis of this property, 
as described in the following sections.

The construction of this article is following.
Section \ref{s3} gives the mathematical formulation concerning
statistical hypothesis testing. 
Section \ref{s2} 
defines the hypothesis scheme for the entanglement of the two-photon
states generated by SPDC. 
Sections \ref{s4} - \ref{s7} describe 
testing methods as well as design of experiment.
Section \ref{s8} examines the experimental aspects of the hypothesis
testing on the entanglement.
The designs on the time allocation are evaluated by the experimental data.

\section{Hypothesis testing for probability distributions}\label{s3}
\subsection{Formulation}
In this section, we review the fundamental knowledge
of hypothesis testing for probability distributions\cite{lehmann}. 
Suppose that a random variable $X$ is 
distributed according to a probability measure $P_\theta$
identified by an unknown parameter $\theta$, 
and that the unknown parameter $\theta$ 
belongs to one of the mutually disjoint sets $\Theta_0$ and $\Theta_1$.
When the task is to guarantee that
the true parameter $\theta$ belongs to the set $\Theta_1$
with a certain significance,
we choose the null hypothesis $H_0$ and 
the alternative hypothesis $H_1$ as
\begin{equation}
	\label{eq:hypo}
	H_0:\theta\in\Theta_0
	\mbox{ vs. }
	H_1:\theta\in\Theta_1.
\end{equation}
The task is then described by a test, where we decide to accept the hypothesis 
$H_1$ by rejecting the null hypothesis $H_0$ with confidence.
We make no decision when we can not reject the null hypothesis.
The test based is characterized as a function
$\phi(x)$ taking a value in $\{0,1\}$; we can reject $H_0$ if $\phi(x) = 1$, 
but can not reject it if $\phi(x) = 0$. 
The test can be also described by the rejection region 
defined by $\{x|\phi(x)=1\}$, whether the data $x$ falls into the rejection
region or not. The function $\phi(x)$ should be designed
properly according to the problem to provide an appropriate test.

For example, suppose that mutually independent data $x_1, \cdots , x_n$ 
obey a normal distribution with unknown mean $\mu$ and 
known variance $\sigma_0^2$
\begin{equation}
      P_{\mu} (x) = \frac{1}{\sqrt{(2\pi \sigma_0^2)^n}} 
      \exp \left[-\frac{\sum_{i=1} ^n (x_i-\mu)^2}{2 \sigma_0^2} \right].
 \label{eq:normal}
\end{equation}
We introduce a hypothesis testing 
\begin{equation}
        H_0:\mu\in\{\mu | \mu \leq \mu_0\}
	\mbox{ vs. }
	H_1:\mu\in\{\mu | \mu > \mu_0\}
 \label{eq:hyponormal}
\end{equation}
to decide the mean is larger than a value 
$\mu_0$ with a confidence level $\alpha$. 
Then the function $\phi $ of the average on the data is defined by
$\bar{x}=(1/n) \sum_{i=1}^{n} x_i$ as
\begin{align}
  \phi(x) =
  \left\{
   \begin{array}{ll}
     &1 , \: \mbox{ if }
     \sqrt{n}(\bar{x}-\mu_0)/ \sigma_0 > z_{\alpha}   \\
     &0 , \:  \mbox{ if }
     \sqrt{n}(\bar{x}-\mu_0)/ \sigma_0 \leq z_{\alpha}, 
  \end{array}
 \right.
 \label{eq:phinormal}
\end{align}
where $z_{\alpha}$ is given by
\begin{align}
   z_{\alpha} &= \Phi ^{-1} (\alpha) \label{eq:za}\\
   \Phi(z) &:= \int _{-\infty} ^z \frac{1}{\sqrt{2\pi}} e^{-\frac{y^2}{2}} dy.
   \label{eq:Phi}
\end{align}
The rejection region is provided as
\begin{equation}
  \{x|\sqrt{n}(\bar{x}-\mu_0)/ \sigma_0 > z_{\alpha} \}
 \label{eq:rr}
\end{equation}
in this test.

\subsection{p-values} \label{pval}
In a usual hypothesis testing, 
we fix our test before applying it to the data.
Sometimes, however, we compare tests in a class $T$
by the minimum risk probability to reject  
the hypothesis $H_0$ with given data.
This probability is called the p-value, which depends on the observed data $x$
as well as the subset $\Theta_0$ to be rejected.
For a given class $T$ of tests, the p-value is defined as
\begin{align}
 pval:=\min_{\phi \in T: \phi(x)=1}
  \max_{\theta \in \Theta_0} P_{\theta} (\phi).
\end{align}
In the hypothesis test for a normal distribution defined previously 
(Eqs. (\ref{eq:normal})-(\ref{eq:Phi})), the p-value is given by
\begin{equation}
   pval=\Phi(\sqrt{n}(\bar{x}-\mu_0)/ \sigma_0).
 \label{eq:pvnormal}
\end{equation}
The concept of p-value is useful
for comparison of several designs of experiment.

\subsection{Likelihood Test}
In mathematical statistics, the likelihood ratio tests 
is often used as a class of standard tests\cite{lehmann}.
Likelihood is a function of unknown parameter $\theta$ defined as
 the probability distribution $P_\theta (x)$ for given data $x$.
%For simplicity,
%$X$ is assumed to be discrete valued.
%Hence, the probability function $p_\theta(x):=P_\theta\{X=x\}$
%is used instead of the probability measure. 
%The same argument for other cases
%is easily obtained by adopting measure theoretic notations.
When both $\Theta_0$ and $\Theta_1$ consist of single elements
as $\Theta_0=\{\theta_0\}$ and $\Theta_1=\{\theta_1\}$,
the rejection region of the likelihood ratio test
%$\phi_{\LR,r}$ 
is defined as
\[
  \left\{ x \left|\frac{P_{\theta_0}(x)}{P_{\theta_1}(x)} < r \right.\right \}.
\]
where $r$ is a constant,
and the ratio
$P_{\theta_0}(x)/P_{\theta_1}(x)$ is called the likelihood ratio.
In the general case, i.e.,
the cases where $\Theta_0$ or $\Theta_1$ has at least two elements,
the likelihood ratio test is given by its rejection region:
\[
\left\{x\left|
	\frac{\sup_{\theta\in\Theta_0}P_\theta(x)}
        {\sup_{\theta\in\Theta_1}P_\theta(x)} <  r 
\right.\right\}.
\]

In the hypothesis test for a normal distribution defined previously 
(Eqs. (\ref{eq:normal})-(\ref{eq:Phi})), logarithm of the likelihood ratio is
given by
\begin{align}
 \log \frac{P_{\mu_0}(x)}{P_{\bar{\mu}=\bar{x}}(x)} 
 &= \frac{1}{2 \sigma_0 ^2} \left\{\sum \left(x_i-\mu_0 \right)^2-
     \sum \left(x_i-\bar{\mu} \right)^2 \right\} \nonumber \\
 &= \lefteqn{- \frac{n(\bar{x}-\mu_0)^2}{2\sigma_0 ^2}}.
\end{align} 
Therefore, the rejection region of the likelihood ratio test coincide with 
Eq. (\ref{eq:rr}), if we set $r=\exp [-z_\alpha^2/2]$.

\section{Hypothesis Testing scheme for entanglement
in SPDC experiments}\label{s2}
\subsection{Formalization}
This section introduces the hypothesis test for entanglement.
We consider the entanglement of two-photon pairs generated by SPDC.
The two-photon state is described by a density matrix $\sigma$.
We assume two-photon generation process to be identical but individual. 
Here we measure the entanglement by the fidelity 
between the generated state $\sigma$ and the maximally entangled target state
$\vert \Phi^{(+)} \rangle $:
\begin{equation}
\fid = \langle \Phi^{(+)} \vert \sigma \vert \Phi^{(+)} \rangle. 
\label{Fidelity}
\end{equation}
The purpose of the test is to guarantee, with a certain significance,
that the state is sufficiently close to the maximally entangled state.
For this purpose, the hypothesis that the fidelity $\fid$ 
is less than a threshold $\fid_0$
should be disproved with a small error probability.
In mathematical statistics, the above situation is 
formulated as hypothesis testing;
 we introduce the null hypothesis $H_0$ that
entanglement is not enough and 
the alternative $H_1$ that the entanglement is enough:
\begin{align}
H_0:\fid \le \fid_0 \hbox{ vs. }
H_1:\fid > \fid_0,
\end{align}
with a threshold $\fid_0$.

The coincidence counts of photon pairs generated in the SPDC experiments
can be assumed to be a random variable
independently distributed according to a Poisson distribution.
In the following, 
a symbol labeled by a pair $(x,y)$ represents
a random variable or parameter related to
the measurement vectors $\ket{x,y}$.
When the dark count is negligible,
the number of detection events (i.e., coincidence counts) $n_{x y}$ 
on the vectors $\ket{x,y}$
is a random variable according to
the Poisson distribution
${\rm Poi}(\lambda \mu_{x y}t_{x y})$
of mean  $\lambda \mu_{x y}t_{x y}$,
where
\begin{itemize}
\item
$\lambda$ is a known constant related to the photon detection rate, 
determined from the averaged photon-pair-generation rate
and the detection efficiency,
\item
$\mu_{x y}=\bra{x,y}\sigma\ket{x,y}$
is an unknown constant,
\item
$t_{x y}$ is a known constant of the time for detection.
\end{itemize}
The probability function of $n_{x y}$ is
\[
  n_{x y}={\rm Poi}(\lambda \mu_{x y}t_{x y})=
	\exp(-\lambda \mu_{x y}t_{x y})
	\frac
	{(\lambda \mu_{x y}t_{x y})^{n_{x y}}}
	{n_{x y}!}
	.
\]
Because the detections at different times are mutually independent,
$n_{x y}$ is independent of $n_{x' y'}$
($x\ne x'$ or $y\ne y'$).
In this paper, we discuss the quantum hypothesis testing 
under the above assumptions,
whereas Usami et al.\cite{usami} 
discussed the state estimation under the same assumptions.

\subsection{Modified visibility}
Visibility of the two-photon interference is an indicator of 
entanglement commonly used in the experiments.
The two-photon interference fringe is obtained 
by the measurement of coincidence counts on the vector $\ket{x,y}$,  
where the vector $\ket{y}$ is rotated along a great circle 
on the Poincare sphere with a fixed vector $\ket{x}$.
The visibility is calculated with the maximum and minimum 
number of coincidence counts, $n_{max}$ and $n_{min}$, as
the ratio $(n_{max}-n_{min})/(n_{max}+n_{min})$.
We need to make the measurement with at least two fixed vectors $\ket{x}$
in order to exclude the possibility of the classical correlation. 
We may choose the two vectors $\ket{H}$ and $\ket{D}$ as $\ket{x}$, 
for example.
However, our decision will contain a bias, 
if we measure the coincidence counts only with two fixed vectors.
The bias in the measurement emerges as the fact that the visibility 
reflects not only the fidelity 
but also the direction of the deviation of the given state 
from the maximally entangled target state.
Hence, we cannot estimate the fidelity 
in a statistically proper way from the visibility. 

In order to remove the bias based on such a direction,
we propose to measure the counts on the coincidence vectors
$|HH\rangle, |VV\rangle, |DD\rangle, |XX\rangle, 
|RL\rangle,$ and $|LR\rangle$,
and the counts of the anti-coincidence vectors
$|HV\rangle, |VH\rangle, |DX\rangle, |XD\rangle, 
|RR\rangle$, and $|LL\rangle$.
The former corresponds to
the maximum coincidence counts in the two-photon interference,
and the latter does to 
the minimum.
Since the equations
\begin{align}
&|HH\rangle\langle HH|+|VV\rangle\langle VV|+
|DD\rangle\langle DD|\nonumber \\
&+|XX\rangle\langle XX|+
|RL\rangle\langle RL|+|LR\rangle\langle LR|\nonumber \\
=&
2|\Phi^{(+)}\rangle\langle\Phi^{(+)}|+ I\label{2-12}
\end{align}
and
\begin{align}
&|HV\rangle\langle HV|+|VH\rangle\langle VH|+|XD\rangle\langle XD|\nonumber \\
&+|DX\rangle\langle DX|+|RR\rangle\langle RR|+|LL\rangle\langle LL|\nonumber \\
=&
2I - 2|\Phi^{(+)}\rangle\langle\Phi^{(+)}|
\end{align}
hold,
In this paper, we call this proposed method the modified visibility method.
Using this method, we can test the fidelity 
between the maximally entangled state 
$|\Phi^{(+)}\rangle \langle \Phi^{(+)}|$
and the given state $\sigma$, using
the total number of counts of the coincidence events 
(the total count on coincidence event)
$n_1$
and 
the total number of counts of the anti-coincidence events 
(the total count on anti-coincidence events)
$n_2$ obtained by measuring on all the vectors
with the time $\frac{t}{12}$.
holds,
we can estimate the fidelity 
by measuring the sum of
the counts of 
the following vectors:
$|HH\rangle, |VV\rangle, |DD\rangle, |XX\rangle, 
|RL\rangle,$ and $|LR\rangle$,
when $\lambda$ is known\cite{BMNMDM03,TMH05}.
This is because
the sum $n_1:= n_{HH}+n_{VV}+ n_{DD}+ n_{XX}+n_{RL}+n_{LR}$ 
obeys the Poisson distribution with the expectation value 
$\lambda \frac{1+ 2\fid }{6}t_1$,
where the measurement time for each vector is $\frac{t_1}{6}$. 
We call these vectors the coincidence vectors because
these correspond to the coincidence events.

However, since the parameter $\lambda$ is usually unknown,
we need to perform another measurement on different vectors 
to obtain
additional information. 
also holds, 
we can estimate the fidelity 
by measuring the sum of
the counts of 
the following vectors:
$|HV\rangle, |VH\rangle, |DX\rangle, |XD\rangle, 
|RR\rangle$, and $|LL\rangle$.
The sum $n_2:= n_{HV}+ n_{VH}+ n_{DX}+ n_{XD}+ n_{RR}+ n_{LL}$
obeys the Poisson distribution 
$\rm Poi(\lambda \frac{2- 2\fid }{6}t_2)$,
where the measurement time for each vector is $\frac{t_2}{6}$. 
Combining the two measurements, we can estimate the fidelity 
without the knowledge of $\lambda$.
We call these vectors the anti-coincidence vectors because
these correspond to the anti-coincidence events.

We can also consider different type of measurement 
on $\lambda$.
If we prepare our device to detect all photons,
the detected number $n_3$ obeys the distribution
$\rm Poi(\lambda t_3$) with the measurement time $t_3$. 
We will refer to it as the total flux measurement.
In the following, we consider the best time allocation for estimation
and test on the fidelity, 
by applying methods of mathematical statistics. 
We will assume that $\lambda$ is known or
estimated from the detected number $n_3$.

\section{Modification of Visibility}\label{s4}
The total count on coincidence events $n_1$ obeys 
Poi$(\lambda\frac{2\fid+1}{12}t)$,
and the count on total anti-coincidence events $n_2$
obeys the distribution Poi$(\lambda\frac{2-2\fid}{12}t)$.
These expectation values $\mu_{1}$ and $\mu_{2}$
are given as 
$\mu_{1} = \lambda\frac{2\fid+1}{12}t$ and 
$\mu_{2} = \lambda\frac{2-2\fid}{12}t$.
Since the ratio
$\frac{\mu_{2}}{\mu_{1}+\mu_{2}}$ is equal to 
$\frac{2}{3}(1-\fid)$,
we can estimate the fidelity using 
the ratio $\frac{n_2}{n_1+n_2}$ as
$\hat{\fid}(n_1,n_2)= 1- \frac{3}{2}\frac{n_2}{n_1+n_2}$.
Its variance is asymptotically equal to 
\begin{align}
\frac{1}
{\lambda(\frac{t}{3(2\fid +1)}+\frac{t}{3(2-2\fid)})}
=
\frac{(2\fid +1)(2-2\fid)}{\lambda t}.
\end{align}
Hence, 
similarly to the visibility, 
we can check the fidelity by using this ratio.

Indeed, when we consider the distribution under the condition 
that the total count $n_1+n_2$ is fixed to $n$,
the random variable $n_2$ obeys the binomial distribution with
the average value $\frac{2}{3}(1-\fid)n$.
Hence, we can apply the likelihood test 
of the binomial distribution.
In this case, by the approximation to the normal distribution,
the likelihood test with the risk probability $\alpha$
is almost equal to the test with the rejection region
concerning the null hypothesis 
$H_0:\fid \le \fid_0$:
$\{(n_1,n_2)|\frac{n_2}{n_1+n_2}\le 
\frac{2}{3}(1-\fid_0)+ 
\Phi^{-1}(\alpha)\sqrt{\frac{(2-2\fid_0)(1+2\fid_0)}{9(n_1+n_2)}}\}$,
where
$\Phi(\alpha):= \int_{-\infty}^\alpha
\frac{1}{\sqrt{2\pi}}e^{-\frac{x^2}{2}}dx$.
The p-value of this kind of tests is 
$\Phi(\frac{n_2(2\fid_{0}+1)- n_1 (2-2\fid_{0})}
{\sqrt{(n_1+n_2)(2\fid_{0}+1)(2-2\fid_{0})}})$.

\section{Design I ($\lambda$: unknown, One-Stage)}\label{s5}
In this section, we consider 
the problem of testing the fidelity 
between the maximally entangled state 
$|\Phi^{(+)}\rangle \langle \Phi^{(+)}|$
and the given state $\sigma$
by performing three kinds of measurement,
coincidence, anti-coincidence, and total flux,
with the times $t_1,t_2$ and $t_3$, respectively.
The data $(n_1,n_2,n_3)$ obeys 
the multinomial Poisson distribution
Poi$(\lambda\frac{2\fid+1}{6}t_1,\lambda\frac{2-2\fid}{6}t_2,\lambda t_3)$
with the assumption that the parameter $\lambda$ is unknown.
In this problem, it is natural to assume that 
we can select the time allocation 
with the constraint for the total time $t_1+t_2+t_3= t$.

As is shown in Hayashi et al \cite{HTM},
when $\fid_1:= 0.899519 \le \fid \le 1$,
the minimum variance of the estimator 
is asymptotically equal to 
$\frac{(1 -\fid)(\sqrt{3}+\sqrt{1-\fid})^2}{\lambda t}$,
which is attained by the optimal time allocation
$t_1=0$, $t_2= \frac{\sqrt{3}}{(\sqrt{3}+\sqrt{1-\fid})}t$,
$t_3 = \frac{\sqrt{1-\fid}}{\sqrt{3}+\sqrt{1-\fid}}t$.
Otherwise, 
it is asymptotically equal to 
$\frac{(2\fid+1)(1-\fid)(\sqrt{2\fid+1}+\sqrt{2-2\fid})^2}{3 \lambda t}$,
which is attained by the optimal time allocation
$t_1= \frac{\sqrt{2-2\fid}}{\sqrt{2\fid+1}+\sqrt{2-2\fid}}t$,
$t_2= \frac{\sqrt{2\fid+1}}{\sqrt{2\fid+1}+\sqrt{2-2\fid}}t$,
$t_3=0$.
This optimal asymptotic variance 
is much better than that obtained 
by the modified visibility method.

Since the optimal allocation depends only on the true parameter $\fid$,
it is suitable to choose the optimal time allocation at the 
threshold $\fid_0$ for 
testing whether the fidelity is greater than
the threshold $\fid_0$.
In the both cases, when we perform the optimal allocation,
the data obeys the binomial Poisson distribution.
Hence, 
similarly to the modified visibility, 
we can check the fidelity by applying 
the likelihood test of binomial distribution to 
the ratio between two kinds of counts.

In the former case, 
the likelihood test with the risk probability $\alpha$
is almost equal to the test with the rejection region
$H_0:\fid \le \fid_0$:
$\{(n_2,n_3)|
\frac{n_2}{n_2+n_3} \le 
\frac{\sqrt{1-\fid_0}}{\sqrt{3}+\sqrt{1-\fid_0}}
+ \frac{\Phi^{-1}(\alpha)}{\sqrt{3}+\sqrt{1-\fid_0}}
\sqrt{\frac{\sqrt{1-\fid_0}\sqrt{3}}
{n_2+n_3}}\}$.
The p-value of this kind of tests is 
$\Phi(
\frac{n_2 \sqrt{3} - n_3 \sqrt{1-\fid_0}}
{\sqrt{(n_2+n_3) \sqrt{1-\fid_0}\sqrt{3}}})$.
In the later case, 
the likelihood test with the risk probability $\alpha$
is almost equal to the test with the rejection region
concerning the null hypothesis 
$H_0:\fid \le \fid_0$:
$\{(n_1,n_2)|
\frac{n_2}{n_2+n_1} \le 
\frac{\sqrt{2-2\fid_0}}{\sqrt{2\fid_0+1}+\sqrt{2-2\fid_0}}
+ \frac{\Phi^{-1}(\alpha)}{\sqrt{2\fid_0+1}+\sqrt{2-2\fid_0}}
\sqrt{\frac{\sqrt{2-2\fid_0}\sqrt{2\fid_0+1}}
{n_1+n_2}}\}$.
The p-value of this kind of tests is 
$\Phi(
\frac{n_2 \sqrt{2\fid_{0}+1}- n_1 \sqrt{2-2\fid_{0}}}
{\sqrt{(n_1+n_2)\sqrt{2\fid_{0}+1}\sqrt{2-2\fid_{0}}}})$.

\section{Design II ($\lambda$: known, One-Stage)}\label{s6}
In this section, we consider the case where $\lambda$ is known.
In this case, 
when $\fid \ge \frac{1}{4}$,
the optimal time allocation is 
$t_2=t$, $t_1=t_3=0$.
That is, 
the count on anti-coincidence 
$(t_1 = 0;t_2 =t)$ is better than
the count on coincidence $(t_1 = t;t_2 = 0)$.
In fact, Barbieri {\it et al.}\cite{BMNMDM03}
measured the sum of 
the counts on the anti-coincidence vectors
$|HV\rangle, |VH\rangle, |DX\rangle, |XD\rangle, |RR\rangle, |LL\rangle$
to realize the entanglement witness in their experiment.
In this case, the count on anti-coincidence $n_2 $
obeys the Poisson distribution with the average
$\lambda \frac{1-\fid}{3}t$,
the fidelity is estimated by $1- \frac{3 n_1}{\lambda t}$.
Its variance is 
$\frac{3 (1-\fid)}{\lambda t}$.
Then, the likelihood test with the risk probability $\alpha$
of the Poisson distribution
is almost equal to the test with the rejection region:
$\{n_2|n_2 \le
\frac{\lambda (1-\fid_0)}{3}t
+ \Phi^{-1}(\alpha)
\sqrt{
\frac{\lambda (1-\fid_0)}{3}t
}\}$ concerning the null hypothesis 
$H_0:\fid \le \fid_0$.
The p-value of likelihood tests 
is 
$\Phi(\frac{n_2- \lambda (1-\fid_0)t/3
}{\sqrt{\lambda (1-\fid_0)t/3}})$.

When $\fid < \frac{1}{4}$,
the optimal time allocation is 
$t_1=t$, $t_2=t_3=0$.
The fidelity is estimated by $\frac{3 n_2}{\lambda t}-\frac{1}{2}$.
Its variance is 
$\frac{3 (1+2\fid) }{2\lambda t}$.
Then, the likelihood test with the risk probability $\alpha$
of the Poisson distribution
is almost equal to the test with the rejection region:
$\{n_1|n_1\ge
\lambda \frac{1+2\fid_0}{6}t
+ \Phi^{-1}(1-\alpha)
\sqrt{
\lambda \frac{1+2\fid_0}{6}t
}\}$
concerning the null hypothesis 
$H_0:\fid \le \fid_0$.
The p-value of likelihood tests is
$\Phi(
\frac{-n_1+\lambda (1+2\fid_0)t/6}
{\sqrt{\lambda (1+2\fid_0)t/6}})$.

\section{Design III ($\lambda$: known, Two-Stage)}\label{s7}
Next, for a further improvement,
we minimize the variance by optimizing the time allocation 
$t_{HV}$, $t_{VH}$, $t_{DX}$, $t_{XD}$, $t_{RR}$, and $t_{LL}$ between the 
anti-coincidence vectors
$B=\{|HV\rangle$, $|VH\rangle$, $|DX\rangle$, $|XD\rangle$, 
$|RR\rangle$, $|LL\rangle \}$, respectively,
under the restriction of the total measurement time:
$t_{HV}+ t_{VH}+ t_{DX}+ t_{XD}+ t_{RR}+ t_{LL}=t$.
The number of the counts $n_{xy}$ obeys Poisson distribution
Poi($\lambda \mu_{xy}t_{xy}$) with the unknown parameter 
$\mu_{xy}=
\langle x_A y_B|\sigma | x_A y_B\rangle$.
the minimum value of estimation of the fidelity 
$\fid= 1- \frac{1}{2}\sum_{(x,y)\in B} \mu_{xy}$
is 
\begin{align}
\frac{
(
%\sqrt{\mu_{HV}}+\sqrt{\mu_{VH}}+\sqrt{\mu_{DX}}+\sqrt{\mu_{XD}}
%+\sqrt{\mu_{RR}}+\sqrt{\mu_{LL}}
\sum_{(x,y)\in B}\sqrt{\mu_{x,y}}
)^2
}{4 \lambda t},\label{2-25-11}
\end{align}
which is attained by the optimal time allocation
\begin{align}
	t_{x y}=
	\frac
	{\sqrt{\mu_{x y}}t}
	{
\sum_{(x,y)\in B}\sqrt{\mu_{x,y}}
%\sqrt{\mu_{HV}}+\sqrt{\mu_{VH}}+\sqrt{\mu_{DX}}+\sqrt{\mu_{XD}}
%+\sqrt{\mu_{RR}}+\sqrt{\mu_{LL}}
},
\end{align}
which is called Neyman allocation and is used in sampling design\cite{Cochran}.

However, this optimal time allocation
is not applicable in the experiment, 
because it depends on the unknown parameters
$\mu_{HV},$ $\mu_{VH},$ $\mu_{DX},$ $\mu_{XD},$ $\mu_{RR},$ and $\mu_{LL}$.
In order to resolve this problem, 
we use a two-stage method, where 
the total measurement time $t$ is divided into
$t_f$ for the first stage and
$t_s$ for the second stage under the condition of $t=t_f+t_s$.
In the first stage,
we measure the counts on the anti-coincidence vectors for $t_f/6$ and
estimate the expectation value for Neyman allocation on measurement time $t_s$.
In the second stage,
we measure the counts on anti-coincidence vectors
$\ket{x_A y_B}$ according to
the estimated Neyman allocation.
The two-stage method is formulated as follows.
\smallskip\\
(i)  The measurement time for each vector
in the first stage is given by
$t_f/6$ 
\\
(ii)
In the second stage,
we measure the counts on a vector 
$\ket{x_A y_B}$ the measurement time $\tilde t_{x y}$
defined as
\[
	\tilde t_{x y}=
	\frac
	{m_{x y}}
	{\sum_{(x,y)\in B}\sqrt{m_{x y}}}
	(t-t_f)
\]
where
$m_{x y}$ is the observed count in the first stage.
\\
(iii)
Define $\hat\mu_{x y}$ and $\hat\fid$ as
\begin{align*}
	\hat{\mu}_{x y}=
	\frac{n_{x y}}{\lambda\tilde t_{x y}},
\quad	\hat\fid=
	1-\frac12\sum_{(x,y)\in B}\hat{\mu}_{x,y},
\end{align*}
where
$n_{x,y}$ is the number of the counts on $\ket{x_A y_B}$
for $\tilde t_{x y}$.
Then, we can estimate the fidelity by $\hat{\fid}$.

We can test our hypothesis by using the likelihood test,
which has the rejection region:
\begin{align*}
\left\{
\vec{n}
\left|
\sup_{\vec{\mu}\cdot \vec{w}\ge c_0}
{\rm Poi}(\vec{\mu})(\vec{n})
\le r 
\sup_{\vec{\mu}\cdot \vec{w}< c_0}
{\rm Poi}(\vec{\mu})(\vec{n})
\right.
\right\},
\end{align*}
where
$w_i:= \frac{1}{2\lambda \tilde{t}_i}$
and $c_0:= 1- \fid_0$.
However, it is very difficult to choose $r$
such that the likelihood test has a given risk probability $\alpha$.
In order to resolve this problem,
Hayashi et al. \cite{HTM} proposed a better method.
For treating this method, 
we number the anti-coincidence vectors by $i=1, \ldots, 6$.
Then, the rejection region with the risk probability $\alpha$ is given by
$\cB_{R_\alpha}$, where
the set $\cB_R$ is defined by 
\begin{align}
\cB_R:=
\left\{\vec{n}'\left|
\sum_{i=1}^6 \frac{n_i'}{\tilde{\mu}_i(R)}
\le 1
\right.\right\},
\end{align}
and $\tilde{\mu}_i(R)$ are defined as follows:
\begin{align}
\frac{c_0}{w_i} - \tilde{\mu}_i(R)+ 
\tilde{\mu}_i(R)\log \frac{\tilde{\mu}_i(R) w_i}{c_0}&= R 
\hbox{~~if }
R\le R_{0,i}
\label{2-6-12}
\\
\frac{c_0}{w_M} + \tilde{\mu}_i(R) \log \frac{w_M-w_i}{w_M}&= R 
\hbox{~~if }
R > R_{0,i}
\label{2-6-11}
\end{align}
where 
$w_M:= \max_i w_i$
and 
$R_{0,i}:=\frac{c_0}{w_M} 
+ \frac{c_0(w_M-w_i)}{w_i w_M}\log \frac{w_M-w_i}{w_M}$.
The value $R_\alpha$ is defined by
\begin{align}
- \min_{i \neq j}z_{i,j}(R_\alpha)= \Phi^{-1}(\alpha),
\end{align}
where $z_{i,j}(R)$ is given by
\begin{align}
z_{i,j}(R):=
\left\{
\begin{array}{ll}
\frac{x_i(R)}{\sqrt{y_i(R)}} & 
\hbox{ if }
\frac{2x_j(R)y_i(R)}{x_i(R)y_i(R)+x_i(R)y_j(R) }\ge 1 
\\
\frac{x_j(R)}{\sqrt{y_j(R)}} & 
\hbox{ if }
\frac{2x_i(R)y_j(R)}{x_j(R)y_j(R)+x_j(R)y_i(R) }\ge 1 .
\end{array}
\right.
\end{align}
Otherwise,
\begin{widetext}
\begin{align}
z_{i,j}(R):=
\frac{2
(x_i(R) x_j(R)(y_i(R) +y_j(R))- x_i(R)^2 y_j(R)- x_j(R)^2 y_i(R))}{
\sqrt{(x_i(R)-x_j(R))(y_i(R)-y_j(R))}
\sqrt{x_i(R) y_j(R)^2+ x_j(R) y_i(R)^2 -y_i(R) y_j(R) (x_i(R)+x_j(R))}
},
\end{align}
\end{widetext}
where
$x_i(R):= \frac{c_0}{w_i \tilde{\mu}_i(R)}-1$
and $y_i(R):= \frac{c_0}{w_i \tilde{\mu}_i(R)^2}$.

Concerning these tests, the p-value is calculated to
\begin{align}
\displaystyle \Phi(
- \min_{i\neq j}
z_{i,j}(R'(\vec{n}))),\label{3-7-2-a}
\end{align}
where $R'(\vec{n})$ is defined as
\begin{align}
\sum_{i=1}^6 \frac{k_i}{\tilde{\mu}_i(R'(\vec{n}))}=1.
\end{align}

\section{Analysis of Experimental Data}\label{s8}
\begin{figure}[tsp]
\begin{center}
\includegraphics[width=8cm]{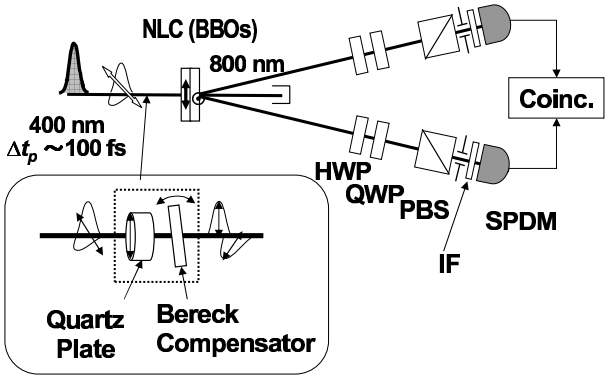}
\end{center}
\caption{Schematic of the entangled photon pair generation by spontaneous parametric down conversion. Cascade of the nonlinear crystals (NLC) generate the photon pairs. Group velocity dispersion and birefringence in the NLCs are pre-compensated with quartz plates and a Bereck compensator. Two-photon states are analyzed with half wave plates (HWP), quarter wave plates (QWP), and polarization beam splitters (PBS). Interference filters (IF) are placed before the single photon counting modules (SPCM).}
\label{SPDC}
\end{figure}

The experimental set-up for the hypothesis testing is shown in Fig. \ref{SPDC}. 
The nonlinear crystals (BBO), the optical axis of which were set to orthogonal to on another, were pumped by a pulsed UV light polarized in $45^{\circ}$  direction to the optical axis of the crystals. One nonlinear crystal generates two photons polarized in the horizontal direction $(\vert HH \rangle)$ from the vertical component of the pump light, and the other generates ones polarized in the vertical direction $(\vert VV \rangle)$ from the horizontal component of the pump.  The second harmonic of the mode-locked Ti:S laser light of about 100 fs duration and 150 mW average power was used to pump the nonlinear crystal. The wavelength of SPDC photons was thus 800 nm. The group velocity dispersion and birefringence in the crystal may differ the space-time position of the generated photons and make the two processes to be distinguished \cite{NUTMK02}. Fortunately, this timing information can be erased by compensation; the horizontal component of the pump pulse should arrive at the nonlinear crystals earlier than the vertical component. The compensation can be done by putting a set of birefringence plates (quartz) and a variable wave-plate before the crystals. We could control 
the two photon state from highly entangled states to separable states by shifting the compensation from the optimal setting. 

\begin{figure}[tsp]
\begin{center}
\includegraphics[width=8cm]{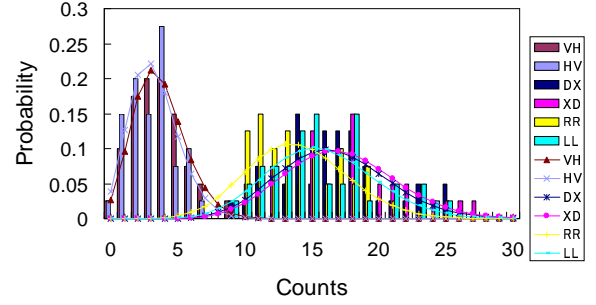}
\end{center}
\caption{Distribution of the counts obtained in one second 
on the vectors 
$\vert VH \rangle,  \vert HV \rangle, \vert XD \rangle, \vert DX \rangle, \vert RR \rangle,$ and $\vert LL \rangle $. Bars present the histograms of 
the measured numbers, and lines show the Poisson distribution with the mean values estimated from the experiment. Measurement time was 40 seconds for each 
vectors.}
\label{coincidence}
\end{figure}

The count on the vector 
$|x_{A}y_{B} \rangle$ was measured by adjusting the half wave plates (HWPs) and the quarter wave plates (QWPs) in  Fig. \ref{SPDC}. 
We accumulated the counts for one second, 
and recorded the counts every one second. 
Therefore, the time allocation of the measurement time on a vector 
must be an integral multiple of one second.
 Figure \ref{coincidence} shows the histogram of the counts in one second on the vector 
\[
B = \{ \vert VH \rangle,  \vert HV \rangle, \vert XD \rangle, \vert DX \rangle, \vert RR \rangle, \vert LL \rangle \},
\]
when the visibility of the two-photon states was estimated to be 0.92. 
The measurement time was 40 seconds on each vector.
The distribution of the coincidence events obeys the Poisson distribution. 
Only small numbers of counts were observed 
on the vectors $\vert HV \rangle$ and $\vert VH \rangle$. 
Those observations agree with the prediction, therefore, 
we expect that the hypothesis testing in the previous sections can be applied. 

In the following, we compare four testing methods on
experimental data with the fixed total time $t$. 
The testing method employ the different time allocations $\{t_{HH}, t_{VV}, t_{DD}, t_{XX}, t_{RL}, t_{LR}, t_{HV}, t_{VH}, t_{DX}, t_{XD}, t_{RR}, t_{LL}\}$ between the measurement vectors:
\begin{description}
\item[(i) Modified visibility method:] $\lambda$ is unknown. The coincidence and the anti-coincidence 
are measured with the equal time allocation;
\begin{align}
&t_{HH}= t_{VV}= t_{DD}= t_{XX}= t_{RL}= t_{LR} \notag \\
&=t_{HV}= t_{VH}= t_{DX}= t_{XD}= t_{RR}= t_{LL} = \frac{t}{12}.
\end{align}
\item[(ii) Design I:] $\lambda$ is unknown. The counts on 
coincidence and anti-coincidence 
are measured with the optimal time allocation 
at the target threshold $\fid_0\le 0.899519$;
\begin{align}
t_{HH}&= t_{VV}= t_{DD}= t_{XX}= t_{RL}= t_{LR}=\frac{t_{1}}{6} \notag\\
t_{HV}&= t_{VH}= t_{DX}= t_{XD}= t_{RR}= t_{LL} = \frac{t_{2}}{6},
\end{align}
where
\begin{align}
t_1&= \frac{t\sqrt{2-2\fid_0}}{\sqrt{2\fid_0+1}+\sqrt{2-2\fid_0}}\notag \\
t_2&= \frac{t\sqrt{2\fid_0+1}}{\sqrt{2\fid_0+1}+\sqrt{2-2\fid_0}}.
\label{t-design1}\end{align} 
\item[(iii) Design II:]  $\lambda$ is known. Only 
the counts on anti-coincidence 
are measured with the equal time allocation
at the target threshold $\fid_0\ge 1/4$;
\begin{align}
t_{HV}&= t_{VH}= t_{DX}= t_{XD}= t_{RR}= t_{LL} = \frac{t}{6}.
\end{align}
\item[(iv) Design III:] $\lambda$ is known. Only the counts 
on anti-coincidence are measured.
The time allocation is given by the two-stage method:
\begin{align} 
t_{HV}= t_{VH}= t_{DX}= t_{XD}= t_{RR}= t_{LL} = \frac{t_{f}}{6}
\end{align}
in the first stage, and
\begin{align}
	\tilde t_{x y}=
	\frac
	{m_{x y}}
	{\sum_{(x,y)\in B}\sqrt{m_{x y}}}
	(t-t_f)
\label{t-design3}
\end{align}
in the second stage. The observed count $m_{x y}$ in the first stage determines 
the time allocation in the second stage. 
\end{description}

We have compared the p-values at the fixed threshold $\fid_0= 7/8 =0.875$ 
with the total measurement time $t=240$ seconds. 
As shown in section \ref{pval}, the p-value measures the minimum risk 
probability to reject the hypothesis $H_{0}$, \textit{i.e.}, 
the probability to make an erroneous decision
to accept insufficiently entangled states with the fidelity less 
than the threshold.
The results of the experiment and the analysis of obtained data are 
described in the following.

In the method (i), we measured the counts on each vectors for $20$ seconds.
We obtained $n_1=9686$ and $n_2=868$ in the experiment, which yielded the p-value $0.343$.

In the method (ii), 
the optimal time allocation was calculated with (\ref{t-design1})
 to be $t_1=55.6$ seconds and $t_2= 184.4$ seconds. 
However, since the time allocation should be the integral multiple of second 
in our experiment, we used the time allocation $t_1= 54$ and $t_2= 186$.
That is, we measure the count on each coincidence vectors for $9$ seconds
and on each anti-coincidence vectors for $31$ seconds.
We obtained $n_1=7239$ and $n_2=2188$ in the experiment, 
which yielded the p-value $0.0715$.

In the method (iii), 
we measured the count on each anti-coincidence vectors for $40$ seconds.
We used $\lambda=290$ estimated from another experiment. 
We obtained $n=2808$ in the experiment, which yielded the p-value $0.0438$.

In the method (iv), the calculation is rather complicated.
Similarly to (iii),
$\lambda$ was estimated to be $290$ from another experiment.
In the first stage, we measured the count on each 
anti-coincidence vectors for $t_{f}/6 = 1$ second.
We obtained the counts $6,3,13,20,11$, and $23$ on the vectors 
$|HV\rangle$, $|VH\rangle$, $|DX\rangle$, $|XD\rangle$, 
$|RR\rangle$, and $|LL\rangle$, respectively.
We made the time allocation of remaining 234 seconds for the second stage according to 
(\ref{t-design3}), and obtained 
$t_{HV}=28.14$, $t_{VH} =19.90$, $t_{DX} =41.42$, $t_{XD} =51.37$, $t_{RR}=38.10$, and
$t_{LL} =55.09$.
Since the time allocation should be the integral multiple of second 
in our experiment, we used the time allocation 
$\{t_{HV}, t_{VH}, t_{DX} , t_{XD} , t_{RR},t_{LL}\}=\{28,20,42,51,38,55\}$.
We obtained the counts on anti-coincidence 
$n_{HV}=99,n_{VH}=66,n_{DX}=703,n_{XD}=863,n_{RR}=531$, and $n_{LL}=853$.
Applying the counts 
and the time allocation  
to the formula (\ref{3-7-2-a}),
we obtained the p-values to be 0.0310.

\begin{figure}[htp]
%\vskip2cm
 \begin{center}
 \includegraphics* [width=8cm]{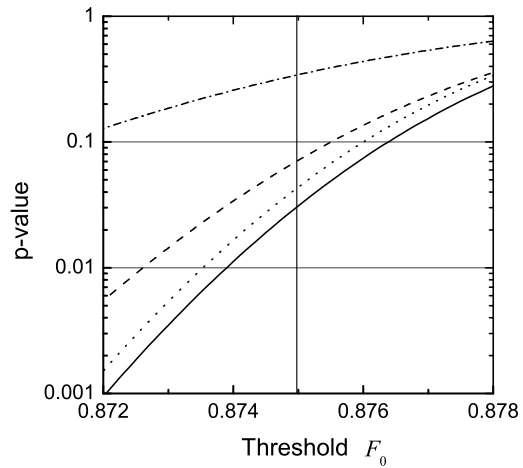}
  \end{center}
   \caption{Calculated p-value as a function of the threshold. 
Dash-dot: (i)the modified visibility, dash: (ii)design I, dot: (iii)design II, 
solid: (iv) design III.}
   \label{fig:relent2}
\end{figure}

The p-values obtained in the four methods are summarized in the table. 
We also calculated the p-values at different values of 
the threshold $\fid_{0}$ as shown in Fig. \ref{fig:relent2}. 
We fixed time allocation for design I at $t_1 = 54$ s and $t_2 = 186$ s.
As clearly seen, the optimal time allocation 
between the coincidence vectors measurement and 
the anti-coincidence vectors measurement reduces 
the risk of a wrong decision on the fidelity (the p-value)
in analyzing the experimental data. 
The counts on the anti-coincidence vectors 
is much more sensitive to the degradation of the entanglement.
This matches our intuition that the deviation from zero provides
 a more efficient measure than that from the maximum does.
 The comparison between (iii) and (iv) shows that the risk
can be reduced further by the time allocation between 
the anti-coincidence vectors, as shown in Fig. \ref{fig:relent2}. 
The optimal (Neyman) allocation implies that the measurement time 
should be allocated preferably to the vectors that yield 
more counts. 
Under the present experimental conditions, the optimal allocation reduces 
the risk probability to  about 75 \%.
The improvement should increased as the fidelity. 
However, the experiment showed almost no gain when the visibility was 
larger than 0.95. 
In such high visibility, errors from the maximally entangled state are 
covered by dark counts, which are independent of the setting of 
the measurement apparatus.

\vspace{1ex}
\begin{center}
\begin{tabular}{|c|c|c|c|c|}
\hline
& (i) & (ii) & (iii) & (iv) \\
\hline
p-value at $0.875$
& 0.343 &0.0715 &0.0438 &0.0310 \\
\hline
%Fisher information at $0.875$\\
%\hline
\end{tabular}
\par
\vspace{2ex}
\end{center}

%\begin{figure}[htbp]
%\vskip2cm
%     \begin{center}
% \includegraphics*[width=8cm]{compare2.eps}
%  \end{center}
%   \caption{Calculated p-value as a function of the threshold (magnified). 
%Dash: (ii)design I, dots: (iii)design II, solid: (iv) design III.}
%   \label{fig:relent3}
%\end{figure}

\section{Conclusion}
We have applied 
the formulation of the hypothesis testing scheme
and the design of experiment for 
the hypothesis testing of entanglement
to the two-photon state generated by SPDC. 
Using this scheme, 
we have handled the fluctuation in the experimental data
properly. It has been experimentally demonstrated 
that the optimal time allocation improves the test in the terms of p-values:
the measurement time should be allocated preferably to the anti-coincidence
vectors in order to reduce the minimum risk probability.
This design is particularly useful for the experimental test, because the
optimal time allocation depends only on the threshold of the test.
We don't need any further information of the probability distribution and 
the tested state.
We have also experimentally demonstrated that 
the test can be further improved by optimizing time allocation 
among the anti-coincidence vectors by using the two-stage method, 
when the error from the maximally 
entangled state is anisotropic.

\end{document}